\documentclass[iop]{emulateapj}
\usepackage{epstopdf}
\RequirePackage{mathrsfs}
\usepackage{lineno}
\usepackage{hyperref}

\newcommand{\water}{H$_2$O}

\newcommand{\um}{$\mu$m}

\newcommand{\teff}{$T_{\mathrm{eff}}$}
\newcommand{\rapsq}{$R_{ap}^2$}
\newcommand{\rchisq}{$\chi^2_{\nu}$}

\slugcomment{Accepted to AJ}

\shorttitle{Prospecting in ultracool dwarfs}
\shortauthors{Mann et al.}

\begin{document}

\title{Prospecting in Ultracool Dwarfs: \\ Measuring the Metallicities of Mid- and Late-M Dwarfs}

\author{Andrew W. Mann\altaffilmark{1,2,3}, Niall R. Deacon\altaffilmark{4}, Eric Gaidos\altaffilmark{5}, Megan Ansdell\altaffilmark{3}, John M. Brewer\altaffilmark{6}, \
Michael C. Liu\altaffilmark{3}, Eugene A. Magnier\altaffilmark{3}, Kimberly M. Aller\altaffilmark{3}}
  
\altaffiltext{1}{Department of Astronomy, University of Texas at Austin} 
\altaffiltext{2}{Harlan J. Smith Fellow}
\altaffiltext{3}{Institute for Astronomy, University of Hawai'i, 2680 Woodlawn Drive, Honolulu, HI 96822} 
\altaffiltext{4}{Max Planck Institute for Astronomy, Konigstuhl 17, Heidelberg, 69117, Germany}
\altaffiltext{5}{Department of Geology \& Geophysics, University of Hawai'i, 1680 East-West Road, Honolulu, HI 96822} 
\altaffiltext{6}{Department of Astronomy, Yale University, New Haven, CT 06511, USA} 


\begin{abstract}
Metallicity is a fundamental parameter that contributes to the physical characteristics of a star. However, the low temperatures and complex molecules present in M dwarf atmospheres make it difficult to measure their metallicities using techniques that have been commonly used for Sun-like stars. Although there has been significant progress in developing empirical methods to measure M dwarf metallicities over the last few years, these techniques have been developed primarily for early- to mid-M dwarfs. We present a method to measure the metallicity of mid- to late-M dwarfs from moderate resolution ($R\sim2000$) $K-$band ($\simeq2.2$~\um) spectra. We calibrate our formula using 44 wide binaries containing an F, G, K, or early M primary of known metallicity and a mid- to late-M dwarf companion. We show that  similar features and techniques used for early M dwarfs are still effective for late-M dwarfs. Our revised calibration is accurate to $\sim0.07$~dex for M4.5--M9.5 dwarfs with $-0.58<$[Fe/H]$<+0.56$ and shows no systematic trends with spectral type, metallicity, or the method used to determine the primary star metallicity. We show that our method gives consistent metallicities for the components of M+M wide binaries. We verify that our new formula works for unresolved binaries by combining spectra of single stars. Lastly, we show that our calibration gives consistent metallicities with the \citet{Mann:2013gf} study for overlapping (M4--M5) stars, establishing that the two calibrations can be used in combination to determine metallicities across the entire M dwarf sequence. 
\end{abstract}

\keywords{stars: abundances --- stars: binaries: visual --- stars: fundamental parameters --- stars: late-type --- techniques: spectroscopic}

\section{Introduction}\label{sec:intro}

M dwarfs have become attractive targets for exoplanet searches \citep[e.g.,][]{Fischer:2012lr}. M dwarfs represent $\sim75\%$ of stars in the solar neighborhood \citep{2006AJ....132.2360H} so their planets weigh heavily on any Galactic planet occurrence calculations. Stellar companions, which can impede giant planet formation \citep{2012ApJ...745...19K}, dilute transits detections, and make Doppler detections more difficult, are less common around M dwarfs than for solar-type stars \citep{Figueira:2012fk}. M dwarf's low masses and small radii enhance Doppler and transit signals, thereby increasing the feasibility of detecting of Earth-sized planets in their habitable zones. These enhancements strengthen considerably from early- to late- M dwarfs. Early M-type dwarfs have masses and radii about half that of the Sun, while late-M dwarfs can have masses and radii $\sim10\%$ that of the Sun \citep{2010ApJ...721.1725D,Boyajian:2012lr} resulting in deeper transit depths and stronger transit signals for an equal size/mass planet. Further, the habitable zone for a late M-type dwarf is 5-10 times closer to the star than for an early M-type dwarf \citep{Kopparapu2013}, resulting in a larger Doppler signal and more likely and frequent transits. 

Studies of M dwarfs have already advanced the study of planet occurrence with stellar mass \citep[e.g.,][]{Johnson:2010lr, Gaidos:2013b} and metallicity \citep{Mann:2012, Mann:2013vn}. Their low masses give additional leverage on any correlation between stellar mass and planet properties, and their large convective zones dilute any metallicity changes from pollution of the photosphere \citep{1997MNRAS.285..403G, 2001ApJ...556L..59P}. However, fully exploiting M dwarfs to advance our knowledge of planet occurrence requires accurate metallicities for the entire sequence of M dwarfs, which are currently unavailable.

The advantages discussed above, among others, have motivated a number of planet surveys specifically targeting mid- to late-M dwarfs, most of which are coming online in the next few years. This includes near-infrared radial velocity surveys like CARMENES \citep{2012SPIE.8446E..0RQ} and the Habitable-Zone Planet Finder \citep{2012SPIE.8446E..1SM}, transit surveys like APACHE \citep{2013EPJWC..4703006S} and MEarth \citep{Nutzman:2008gf, Charbonneau:2009rt}, and direct imaging searches like PALMS \citep{2012ApJ...753..142B}. Some of these surveys are directed at M dwarfs generally, but many are aimed at mid- to late-M dwarfs specifically. The Habitable-Zone Planet Finder, for example, is targeting M4-M9 dwarfs \citep{2012SPIE.8446E..1SM}.

Our knowledge of planet parameters are directly linked to our understanding of their host stars. Thus, these surveys require reliable stellar masses, radii, and metallicities to properly characterize orbiting planets that are discovered. The {\it Gaia} spacecraft \citep{2012Ap&SS.341...31D} is expected to measure parallaxes for the majority of the M dwarfs targeted by these surveys \citep{Bailer-Jones:2013aa}, and distances can be used to derive luminosities and infer masses and radii \citep[e.g.,][]{2000A&A...364..217D, 2006ApJ...651.1155B}, but not metallicities. M dwarfs have sufficiently cool atmospheres to enable the formation of molecules with complex absorption bands. These bands are difficult to model but dominate the visible spectrum, making continuum identification difficult and creating line confusion. The result is that model-dependent methods such as spectral synthesis and curve of growth analysis that work well on solar-type stars are ineffective for M dwarfs \citep[although improvements are ongoing, e.g.,][]{Onehag:2012lr}.

An alternative approach is to measure the metallicity of M dwarfs using empirical techniques. Such methods include measuring the position on a color-magnitude diagram \citep[e.g.,][]{Schlaufman:2010qy, 2012A&A...538A..25N}, the strength of molecular lines in the optical \citep[e.g.,][]{Woolf:2006uq, Dhital:2012lr}, or atomic lines in the optical or near-infrared \citep[e.g.,][]{2010ApJ...720L.113R,Mann:2013gf}. Colors such as $g-r$ \citep[e.g.,][]{West:2004qy, Bochanski:2013lr} or $JHK$ \citep[e.g.,][]{Johnson:2012fk,Newton:2014} can be used for predicting metallicity, but the errors are higher than other approaches and are subject to additional systematic errors \citep{Mann:2012}. These methods are typically calibrated using wide binaries containing a solar-type primary \citep[e.g.,][]{2005A&A...442..635B}. This assumes that wide binaries formed from the same molecular cloud and therefore have the same metallicity, which is well-established for solar-mass binaries \citep{2004A&A...420..683D,2006A&A...454..581D}.

To date, calibrations of these empirical techniques have been created only for early and mid M-type dwarfs. The calibration from \citet[][henceforth M13]{Mann:2013gf} utilized the largest sample, but it contained only one M5, one M6, and nothing later. \citet{Newton:2014} focused on cooler M dwarfs for the MEarth survey, but included no M6 dwarfs, a single M7, and nothing cooler. As a result, the effectiveness of these calibrations for the coolest M dwarfs remains untested.

The faintness at optical wavelengths of the coolest M dwarfs makes them less likely to show up in long-time baseline proper motion surveys, because they generally rely on detections in the optical. Thus, until recently, it was difficult to locate wide, common proper motion pairs containing a late-type M dwarf.  However, the problem has been mitigated significantly thanks to wide-field digital sky surveys such as the Sloan Digital Sky Survey \citep{Aihara:2011ly, West:2011fj}, the Two-Micron All-Sky Survey \citep[2MASS][]{Skrutskie:2006lr}, and the Panoramic Survey Telescope and Rapid Response System \citep[PAN-STARRS][]{2010SPIE.7733E..12K}, which have provided proper motions and photometry for cooler and fainter objects than were previously accessible. Furthermore, now that methods to estimate the metallicity of early-M types are established, we can use pairs of early-M and late-M type dwarfs to extend the calibration to cooler temperatures.

In this paper we investigate methods to measure the metallicities of M4.5--M9.5 dwarfs. We use 44 wide binaries containing an F, G, K, or early M dwarf primary and an M4.5--M9.5 companion. We determine metallicities for the primary stars by combining those from the literature with our own observations. Following the techniques of M13, we derive empirical calibrations between features in $K-$band spectra and the metallicity of ultracool dwarfs. In Section~\ref{sec:sample} we present our wide binary sample. In Section~\ref{sec:obs} we describe our observations of the primary and companion stars. In Section~\ref{sec:analysis} we detail our calculations of metallicities for the primary stars and spectral classifications of the companions. We test how prior metallicity calibrations work on our ultracool dwarf sample in Section~\ref{sec:priorcal} then derive a new calibration in Section~\ref{sec:newcal}. We investigate the reliability of the calibration by employing a number of tests in Section~\ref{sec:tests}. We conclude with a brief summary of our work in Section~\ref{sec:discussion}.

All wavelengths used in this work are stated as vacuum values.

\section{Sample}\label{sec:sample}
We constructed our wide binary sample from literature sources and from our own analysis of proper motion catalogs. Our sample contains M dwarfs as early as M4.5 for overlap with prior studies, but also includes stars out to the end of the M dwarf sequence at M9.5 (see Section~\ref{sec:spt} for details on spectral types are assigned). We selected pairs with $-30^\circ<\delta<68^\circ$ (easily reachable from Mauna Kea telescopes), companions with $K<13$ (reasonable integration times), and pairs with primary-companion separation $>5\arcsec$ (tighter binaries are difficult to observe and may have contaminated photometry). We also required that the primaries have spectral types later than F6, as stars earlier than this often have very few lines useful for metallicity determination. It is possible to determine the metallicities of M dwarfs as late as $\simeq$M5 using empirical techniques, although these methods are best tested for stars M4 and earlier, so we conservatively restrict our primary stars to F6 to M3. 

In total we identified 61 pairs meeting the above criteria.
\begin{itemize}
\item 52 are benchmark systems previously identified in the literature (see Table~\ref{tab:sample} for list of references). 
\item Twelve pairs were taken from an ongoing search for ultracool and brown dwarf companions to {\it Hipparcos} stars using astrometry from the 2MASS and Pan-STARRS1 \citep{2012ApJ...757..100D, DeaconPrep}. 
\item Thirteen more targets were found by looking for co-moving pairs in \citet{2005AJ....129.1483L} and \citet[][LG11]{Lepine:2011vn} proper motion catalogs following the methods outlined in \citet{Lepine:2007qy}. Although this matching returned many targets already identified in the literature, this included seven early M + late-M pairs which were previously overlooked. 
\item A single final target was discovered in the finder while observing the primary (J19074+5905) for a separate project. We obtained a low-S/N spectrum of the companion, which enabled us to measure the spectral type and confirm the primary and companion have consistent distances based on the $M_K$-spectral type relation from \citet{Lepine:2013lr}.
\end{itemize}
Note that many targets are found in multiple sources.

We observed all 61 targets (Section~\ref{sec:obs}). However, 17 of them were rejected from the final sample because: (1) the companion was earlier than M4.5 or an L dwarf, (2) the S/N of the observations was too low ($\lesssim60$) to be useful, (3) the primary was too hot or cold to derive an accurate metallicity and had no reliable metallicity from the literature, or (4) the primary was a double-lined spectroscopic binary, which complicates the analysis (see Section~\ref{sec:primary} for more details).

\begin{figure}[htbp] 
   \centering
   \includegraphics[width=0.5\textwidth]{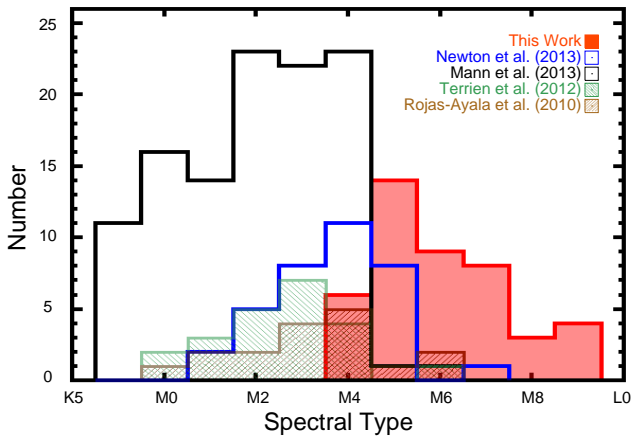} 
   \caption{Distribution of spectral types covered by our calibration sample (red) and calibrators used in prior studies of M dwarf metallicities \citep{Rojas-Ayala:2012uq, Terrien:2012lr, Mann:2013gf,Newton:2014}. We consider only reported spectral types, and do not account for systematic differences between spectral typing methods. Note that many sources contain overlapping targets.}
   \label{fig:sample}
\end{figure}

In Figure~\ref{fig:sample} we show the distribution of spectral types for companion stars in this study as well as those from prior studies of M dwarf metallicities. Although there is significant overlap in the M4.5--M5.5 range, we have greatly expanding the number of companions with spectral types M6 and later. The full sample of binaries used for calibration is listed Table~\ref{tab:sample}, including the spectral types (Section~\ref{sec:spt}), primary star metallicities (Section~\ref{sec:primary}) and references establishing the binarity of the pair. Because naming conventions for these stars vary throughout the literature, we also provide the coordinates of each target. 

\begin{deluxetable*}{l l l l | l l l l l l }
\caption{Wide Binary Sample}
\tablewidth{0pt}
\tablehead{
\multicolumn{4}{c}{Companion} &  \colhead{} & \multicolumn{5}{c}{Primary} \\
\cline{1-4}  \cline{6-10} \\
\colhead{Name} & \colhead{R.A.}& \colhead{$\delta$} & \colhead{SpT} & \colhead{} & \colhead{Name} & \colhead{SpT} & \colhead{[Fe/H]$^a$} & \colhead{[Fe/H] Ref$^b$} & \colhead{Binary Ref$^c$} \\
}
\startdata
HIP 70623 B & 14$^h$26$^m$45$^s$.74 & -5$^\circ$10\arcmin20\arcsec & M4.5 &  & HIP 70623 & K0V & $+0.56$ $\pm$ 0.03 & SPOCS & 15,18\\
LSPM J0212+1249W & 2$^h$12$^m$19$^s$.76 & +12$^\circ$49\arcmin25\arcsec & M4.5 &  & NLTT 7300 & M3V & $+0.46$ $\pm$ 0.10 & SpeX & 17\\
HIP 98535 B & 20$^h$1$^m$2$^s$.17 & +48$^\circ$16\arcmin27\arcsec & M4.5 &  & HIP 98535 & G5V & $-0.18$ $\pm$ 0.03 & SME & 15,18\\
2M 1743+2136 & 17$^h$43$^m$15$^s$.32 & +21$^\circ$36\arcmin4\arcsec & M4.5 &  & HIP 86722 & K0V & $-0.39$ $\pm$ 0.05 & F08 & 11,15\\
LSPM J2047+1051N & 20$^h$47$^m$16$^s$.75 & +10$^\circ$51\arcmin45\arcsec & M4.5 &  & HIP 102582 & K2V & $-0.53$ $\pm$ 0.03 & SME & 15\\
NLTT 15867 & 5$^h$58$^m$17$^s$.18 & -4$^\circ$38\arcmin1\arcsec & M4.5 &  & HIP 28267 & G7V & $-0.10$ $\pm$ 0.03 & SPOCS & 5\\
PM I14254+2035 & 14$^h$25$^m$25$^s$.89 & +20$^\circ$35\arcmin45\arcsec & M5.0 &  & HIP 70520 & F9V & $-0.57$ $\pm$ 0.05 & R07 & 9\\
NLTT 8870 & 2$^h$45$^m$41$^s$.23 & +44$^\circ$57\arcmin2\arcsec & M5.0 &  & HIP 12886 & M1V & $+0.11$ $\pm$ 0.10 & SpeX & 5\\
HIP 114424 B & 23$^h$10$^m$22$^s$.08 & -7$^\circ$48\arcmin54\arcsec & M5.0 &  & HIP 114424 & K0V & $+0.10$ $\pm$ 0.03 & SPOCS & 15\\
HIP 114456 B & 23$^h$10$^m$54$^s$.78 & +45$^\circ$30\arcmin43\arcsec & M5.0 &  & HIP 114456 & K0V & $+0.21$ $\pm$ 0.03 & SPOCS & 15,18\\
LSPM J0253+6321 & 2$^h$53$^m$15$^s$.55 & +63$^\circ$21\arcmin6\arcsec & M5.0 &  & HIP 13394 & G0V & $-0.14$ $\pm$ 0.08 & C11 & 9\\
LSPM J1841+2447N & 18$^h$41$^m$9$^s$.81 & +24$^\circ$47\arcmin19\arcsec & M5.0 &  & GJ 1230A & M3V & $+0.18$ $\pm$ 0.10 & SpeX & 17\\
HIP 106551 B & 21$^h$34$^m$45$^s$.17 & +38$^\circ$31\arcmin0\arcsec & M5.0 &  & HIP 106551 & K1III & $+0.05$ $\pm$ 0.06 & C01 & 15\\
LSPM J0932+2659E & 9$^h$32$^m$48$^s$.25 & +26$^\circ$59\arcmin43\arcsec & M5.5 &  & HIP 46843 & G9V & $-0.09$ $\pm$ 0.05 & F08 & 4,5,9,13\\
LSPM J1659+0635 & 16$^h$59$^m$5$^s$.58 & +6$^\circ$35\arcmin32\arcsec & M5.5 &  & HIP 83120 & K0V & $+0.25$ $\pm$ 0.03 & SME & 9,5\\
LSPM J1207+1302 & 12$^h$7$^m$24$^s$.01 & +13$^\circ$2\arcmin13\arcsec & M5.5 &  & HIP 59126 & K0V & $-0.02$ $\pm$ 0.03 & SME & 4,5,9\\
PM I19074+5905 B & 19$^h$7$^m$24$^s$.83 & +59$^\circ$5\arcmin9\arcsec & M5.5 &  & I19074+5905 & M2V & $+0.30$ $\pm$ 0.10 & SpeX & 16\\
I10005+2717 & 10$^h$0$^m$35$^s$.71 & +27$^\circ$17\arcmin6\arcsec & M5.5 &  & HIP 49046 & M1V & $+0.26$ $\pm$ 0.10 & SpeX & 13,15,18\\
LSPM J1748+1143 & 17$^h$48$^m$44$^s$.32 & +11$^\circ$43\arcmin47\arcsec & M5.5 &  & HIP 87182 & K4V & $+0.02$ $\pm$ 0.03 & SME & 9\\
LSPM J0731+1958 & 7$^h$31$^m$38$^s$.88 & +19$^\circ$58\arcmin32\arcsec & M5.5 &  & HIP 36607 & K0V & $+0.05$ $\pm$ 0.03 & SME & 4,5,9,13\\
LSPM J1302+3227 & 13$^h$2$^m$20$^s$.81 & +32$^\circ$27\arcmin10\arcsec & M5.5 &  & HIP 63636 & G8IV & $+0.05$ $\pm$ 0.04 & T05 & 9\\
LSPM J1124+2330E & 11$^h$24$^m$40$^s$.17 & +23$^\circ$30\arcmin57\arcsec & M5.5 &  & NLTT 27298 & M2V & $+0.08$ $\pm$ 0.10 & SpeX & 17\\
LSPM J2049+3216W & 20$^h$49$^m$13$^s$.75 & +32$^\circ$16\arcmin51\arcsec & M6.0 &  & HIP 102766 & K2V & $-0.02$ $\pm$ 0.03 & SME & 5,9\\
PM I10008+3155 & 10$^h$0$^m$50$^s$.19 & +31$^\circ$55\arcmin44\arcsec & M6.0 &  & HIP 49081 & G3V & $+0.20$ $\pm$ 0.03 & SPOCS & 14\\
NLTT19472 & 8$^h$24$^m$52$^s$.44 & -3$^\circ$41\arcmin1\arcsec & M6.0 &  & HIP 41211 & F8V & $-0.28$ $\pm$ 0.08 & C11 & 5\\
NLTT28453 & 11$^h$45$^m$35$^s$.39 & -20$^\circ$21\arcmin4\arcsec & M6.0 &  & HIP 57361 & M2V & $-0.05$ $\pm$ 0.10 & SpeX & 5\\
LSPM J1210+1858E & 12$^h$10$^m$9$^s$.79 & +18$^\circ$58\arcmin7\arcsec & M6.5 &  & HIP 59310 & K3V & $+0.30$ $\pm$ 0.03 & SME & 13,15,18\\
HIP 81910 B & 16$^h$43$^m$49$^s$.50 & -26$^\circ$48\arcmin40\arcsec & M6.5 &  & HIP 81910 & G3V & $+0.24$ $\pm$ 0.03 & SPOCS & 15\\
LSPM J0942+2351 & 9$^h$42$^m$57$^s$.18 & +23$^\circ$51\arcmin19\arcsec & M6.5 &  & NLTT 22411 & M1V & $+0.05$ $\pm$ 0.10 & SpeX & 17\\
PM I11055+4331 & 11$^h$5$^m$30$^s$.90 & +43$^\circ$31\arcmin17\arcsec & M6.5 &  & HIP 54211 & M2V & $-0.32$ $\pm$ 0.10 & SpeX & 5\\
2M 0318+0828 & 3$^h$18$^m$42$^s$.14 & +8$^\circ$28\arcmin0\arcsec & M7.0 &  & NLTT 10534 & M2V & $+0.19$ $\pm$ 0.10 & SpeX & 12\\
PM I16555-0823 & 16$^h$55$^m$35$^s$.29 & -8$^\circ$23\arcmin40\arcsec & M7.0 &  & HIP 82817 & M3V & $-0.08$ $\pm$ 0.10 & SpeX & 13\\
2M 1320+0957 & 13$^h$20$^m$41$^s$.59 & +9$^\circ$57\arcmin50\arcsec & M7.0 &  & HIP 65133 & K4V & $+0.07$ $\pm$ 0.03 & SME & 10\\
NLTT 36549 & 14$^h$12$^m$12$^s$.13 & -0$^\circ$35\arcmin16\arcsec & M7.5 &  & NLTT 36548 & M3V & $-0.26$ $\pm$ 0.10 & SpeX & 4\\
2M 1200+2048 & 12$^h$0$^m$32$^s$.92 & +20$^\circ$48\arcmin51\arcsec & M7.5 &  & G 121-42 & M2V & $-0.15$ $\pm$ 0.10 & SpeX & 10\\
GJ 569B & 14$^h$54$^m$29$^s$.36 & +16$^\circ$6\arcmin8\arcsec & M7.5 &  & HIP 72944 & M2V & $-0.08$ $\pm$ 0.10 & SpeX & 3\\
2M 1916+0509 & 19$^h$16$^m$57$^s$.60 & +5$^\circ$9\arcmin1\arcsec & M7.5 &  & HIP 94761 & M2V & $+0.11$ $\pm$ 0.10 & SpeX & 1\\
2M 2331-0406 & 23$^h$31$^m$1$^s$.64 & -4$^\circ$6\arcmin19\arcsec & M8.0 &  & HIP 116106 & F8V & $-0.26$ $\pm$ 0.03 & SPOCS & 7\\
2M 0003-2822 & 0$^h$3$^m$42$^s$.28 & -28$^\circ$22\arcmin41\arcsec & M8.0 &  & HIP 296 & G8V & $+0.31$ $\pm$ 0.03 & SME & 3,7\\
2M 0430-0849 & 4$^h$30$^m$51$^s$.57 & -8$^\circ$49\arcmin0\arcsec & M8.5 &  & LP 655-23 & M3V & $+0.03$ $\pm$ 0.10 & SpeX & 7\\
HIP 78184 B & 15$^h$57$^m$55$^s$.32 & +59$^\circ$14\arcmin25\arcsec & M9.0 &  & HIP 78184 & M0V & $+0.08$ $\pm$ 0.10 & SpeX & 6,15,18\\
2M 2010+0634 & 20$^h$10$^m$35$^s$.39 & +6$^\circ$34\arcmin36\arcsec & M9.0 &  & NLTT 48838 & M3V & $-0.01$ $\pm$ 0.10 & SpeX & 12\\
2M 0739+1305 & 7$^h$39$^m$43$^s$.85 & +13$^\circ$5\arcmin6\arcsec & M9.0 &  & BD+13 1727 & K2V & $+0.15$ $\pm$ 0.03 & SME & 5\\
2M 2237+3922 & 22$^h$37$^m$32$^s$.55 & +39$^\circ$22\arcmin39\arcsec & M9.5 &  & HIP 111685 & M1V & $+0.03$ $\pm$ 0.10 & SpeX & 2\\
\enddata
\tablenotetext{a}{[Fe/H] values shown here include our applied corrections (see Section~\ref{sec:analysis}).}
\tablenotetext{b}{Metallicity References -- C01 = \citet{2001A&A...373..159C}, SPOCS = \citet{2005ApJS..159..141V}, T05 = \citet{2005PASJ...57...27T}, R07 = \citet{2007A&A...465..271R}, F08 = \citet{2008MNRAS.384..173F}, C11 = \citet{2011A&A...530A.138C}, SME = Spectroscopy Made Easy analysis of CFHT/ESPaDOnS data, SpeX = empirical calibrations of M13 applied to IRTF/SpeX data}
\tablenotetext{c}{Binary Reference --  1 = \citet{1944AJ.....51...61V},  2 = \citet{2001PASP..113..814K},  3 = \citet{2000ApJ...529L..37M},  4 = \citet{Chaname:2004lr},  5 = \citet{Gould:2004fk},  6 = \citet{2006MNRAS.368.1281P},  7 = \citet{2007ApJ...667..520C},  8 = \citet{2007AJ....133..439C},  9 = \citet{Lepine:2007qy}, 10 = \citet{Faherty:2010}, 11 = \citet{2012AJ....144...62A}, 12 = \citet{2012ApJ...760..152L}, 13 = \citet{Tokovinin:2012fj}, 14 = \citet{Mann:2013gf}, 15 = \citet{DeaconPrep}, 16 = SpeX Finder, 17 = \citet{2005AJ....129.1483L} + \citet{Lepine:2011vn}, 18 = Washington Double Star Catalog.}
\label{tab:sample}
\end{deluxetable*}

\section{Observations and Reduction}\label{sec:obs}
\subsection{ESPaDOnS/CFHT}\label{sec:cfhtobs}
We observed 15 F-, G-, and K-type primary stars with the Canada-France-Hawaii Telescope (CFHT) Echelle SpectroPolarimetric Device for the Observation of Stars \citep[ESPaDOnS;][]{Donati:2003uq}. Observations were taken in queued service mode, in the star+sky setting on ESPaDOnS. This yielded a resolution of $\lambda/\Delta\lambda \simeq 65000$ and wavelength coverage from 0.37~\um to 1.05~\um. Because of cosmic rays and atmospheric variations we did not use exposure times higher than 2400s. For fainter sources, we took multiple exposures and stacked them after reduction. All final spectra had S/N $>100$ at 0.67~\um, and typical S/N was $>150$ per resolving element. The data were reduced automatically by the \texttt{Libre-ESpRIT} pipeline described in \citet{Donati:1997fj}. Four of the 15 stars with ESPaDOnS spectra were later rejected because of complications with the primary (see Section~\ref{sec:sample} for more details).

\subsection{SpeX/IRTF}\label{sec:spex}
We obtained near-infrared spectra of the 61 companions and 19 M dwarf primaries with the SpeX spectrograph \citep{Rayner:2003lr} attached to the NASA Infrared Telescope Facility (IRTF) on Mauna Kea. We used the cross-dispersed mode with the 0.3$\arcsec$ slit. This provided simultaneous coverage from 0.8 to 2.4~\um\ at a resolution of $R\simeq2000$. Targets were placed at two positions along the slit (A and B). We took exposures following an ABBA slit-nodding pattern, with at least six exposures per target. Although spectral features used for calculating metallicities (see Section~\ref{sec:newcal}) are free of telluric and OH lines, we still choose to minimize the effect of \water{} and other atmospheric variation by capping individual exposure times were at 120~s. 

Resulting S/N in the $H$ and $K$ bands for all spectra was $>60$ (typically $>90$) for the companions, and $>100$ (typically $>120$) for the early-M dwarf primaries. To avoid effects of flexure in the optical path, we obtained flat-field and argon lamp calibration data at or near the same pointing as the target. We observed an A0V-type star within 1 hr and 0.1 airmasses of each target to remove telluric lines. The faintest targets took more than 1 hr of time (including overhead), and many targets moved by more than 0.1 airmasses over the course of an observation sequence. In these cases we took two A0V stars, sometimes slewing away from the target between exposures to observe one of the A0V stars, then slewing back to take additional target exposures. 

We extracted and reduced spectra using the \texttt{Spextool} IDL package \citep{Cushing:2004fk}. \texttt{Spextool} performed flat-field correction, wavelength calibration, sky subtraction, as well as extraction of the 1D spectrum. We stacked multiple exposures with the IDL routine \texttt{xcombspec} (part of \texttt{Spextool}). While running \texttt{xcombspec} we checked (by eye) to see if exposures were consistent with each other and to remove outliers. However, we removed only three images, and all were taken through thick clouds ($>2$ magnitudes of extinction) and had relatively low S/N. After the spectra were stacked, we performed telluric corrections and flux calibration using the A0V stars with the {\texttt{xtellcor} package \citep{Vacca:2003qy}. Separate orders were combined with the \texttt{xmergeorders} IDL routine. 

As a test, we tried performing telluric correction using different A0V stars for a given target taken in the same night. This generated small color differences amounting to $J-Ks<0.04$, which is consistent with those found by \citet{2013ApJ...779..188M}. The change in overall shape is likely due to seeing changes between the target and standard star \citep{Rayner:2009kx}. Such color terms were only significant when measured across the whole $JHK$ spectrum, and thus were unlikely to effect our results. Within each echellette order, we found spectra were consistent within errors, provided the A0V standard was taken within 0.15 airmasses. 

Reduced spectra were put in vacuum wavelengths using the formula from \citet{Ciddor:96}. We put spectra in the stars' rest frame by cross-correlating them with the spectra of template stars from the IRTF spectral library \citep{Cushing:2005lr, Rayner:2009kx}}. For stars of M4.5 to M6.5 we used the M5V template Gl 51, and for later-type stars we used the M9V template LHS 2065. As a test we tried cross-correlating our spectra with different templates from the IRTF library and found that differences in the derived radial velocity offset were small ($\sim 1$ resolving element) provided the template was within $\simeq$3 spectral subtypes of the target star. 

\citet{Newton:2014} noted that for high S/N spectra \texttt{Spextool} underestimates the error, because of the presence of correlated noise. However, \citet{Newton:2014} targets are significantly brighter and have higher S/N than our targets (S/N $>200$ versus $>90$). At S/N $>200$ Poisson errors are almost negligible, but are likely the dominant source of noise for our spectra. As a test, we took the individual 1D spectra (just prior to stacking with \texttt{xcombxpec}) of 15 random stars from our sample and then re-stacked them into two spectra of each star (using just half the unstacked spectra in each case). We did the same for the corresponding A0V star. We find that differences in equivalent widths between the stacks of the same star are consistent within Poisson or photon noise-based errors except the five stars where the S/N was $\gg$ 150. For these targets the errors are scaled according to the scatter in the stacking.

Reduced spectra of all companions are included with this manuscript.

\section{Analysis}\label{sec:analysis}
\subsection{Primary Star Metallicities}\label{sec:primary}
As with M13, we drew metallicities for the primaries both from the literature and from our own observations. Seven primaries have metallicities in the Spectroscopic Properties of Cool Stars catalog \citep[SPOCS,][]{2005ApJS..159..141V}, which is based on analysis of high-resolution spectra with the Spectroscopy Made Easy \citep[SME,][]{1996A&AS..118..595V} software package. Another seven primaries have metallicities from various literature sources, most of which make use of high-resolution spectra and MOOG \citep{1973PhDT.......180S}. 

Different literature sources use slightly different techniques and thus may have small systematic inconsistencies. As in M13, we corrected for this by using stars common to both the given literature source and the SPOCS sample. Literature sources with less than 30 stars of overlap with SPOCS were not utilized. Generally the overlap sample is $\ll100$ stars, and the corrections are $\ll0.1$~dex. This method enabled us to put all metallicities on the same scale (in this case the SPOCS scale) and estimate the error from residual scatter after applying the correction. More details on the corrections, number of overlapping stars, and derived errors can be found in M13.

We observed 15 of the FGK primaries with CFHT/ESPaDOnS, although four of these stars were removed because of complications with their primary (see Section~\ref{sec:sample}). To determine stellar properties for these stars we modeled each spectrum with the SME software \citep{1996A&AS..118..595V}, fitting to the set of lines tuned for the SPOCS catalog \citep{2005ApJS..159..141V}.  We simultaneously solved for surface gravity, effective temperature, projected rotational velocity, and individual abundances of Na, Si, Ti, Fe, and Ni as in the SPOCS analysis. Solar values were assumed for all of the initial models and after obtaining an initial fit, we perturbed \teff by $\pm 100$K and fit again. Corrections based on Vesta and stellar binary observations as detailed in \citet{2005ApJS..159..141V} were then applied. The SME-determined [Si/Fe] was used as a proxy for alpha-element enhancement. 

\citet{2012ApJ...757..161T} showed that different analysis methods of high-resolution spectra can produce systematically different results for the same stars and that fitting for all parameters simultaneously can create strong correlations between [Fe/H], log~$g$, and \teff. We mitigated this effect by utilizing {\it Hipparcos} parallaxes \citep[][where available]{van-Leeuwen:2005kx,van-Leeuwen:2007yq} and the Yonsei--Yale evolutionary models \citep{2004ApJS..155..667D} to independently constrain log~$g$, following the method from \citet{Valenti:2009fk}. We first used the distance and color to derive a bolometric luminosity. We combined this with the \teff{}, [Fe/H], and [Si/Fe] from an initial fit using SME, which we interpolated onto the Yonsei--Yale grid to get log~$g$. This log~$g$ was compared to the value determined by SME, and if the two did not match, the SME analysis was run again with the gravity fixed to the isochrone value. The process was repeated until the log~$g$ values agreed. The iterative process did not converge for a single star (HIP 102582), which is probably due to an erroneous parallax, color, unresolved companion (tertiary), and/or the relatively low temperature of this star (\teff$\simeq$ 4600~K). However, because the [Fe/H] did not significantly change during the iteration, the metallicity for this object is likely reliable and included in our sample. Final stellar parameters (T$_{eff}$, log~$g$, [M/H], etc.) for stars observed with ESPaDOnS are listed in Table~\ref{tab:primaries}. 

 \begin{deluxetable*}{l  cccccll}
 \tablecaption{Parameters of Primary Stars Observed at CFHT}
 \tablehead{
 \colhead{Name} & \colhead{\teff\tablenotemark{a}}& \colhead{log~$g$\tablenotemark{a}} & \colhead{[Fe/H]\tablenotemark{a}} & \colhead{[M/H]~$\pm~\sigma$} & \colhead{[Na/H]~$\pm~\sigma$} & \colhead{Run Type\tablenotemark{b}} \\
 }
 \startdata
HIP 296 & 5561 &  4.50 & $+0.31$ & $+0.24~\pm$  0.03 & $+0.32~\pm$  0.03 & ITER \\
HIP 36607 & 5077 &  4.55 & $+0.05$ & $+0.02~\pm$  0.07 & $+0.10~\pm$  0.04 & ITER \\
BD+13 1727 & 5098 &  4.70 & $+0.15$ & $+0.10~\pm$  0.04 & $+0.11~\pm$  0.03 & SME VESTA \\
HIP 59126 & 4862 &  4.61 & $-0.02$ & $-0.05~\pm$  0.04 & $-0.00~\pm$  0.04 & ITER \\
HIP 59310 & 4739 &  4.60 & $+0.30$ & $+0.25~\pm$  0.03 & $+0.43~\pm$  0.04 & ITER \\
HIP 65133 & 4604 &  4.64 & $+0.07$ & $+0.00~\pm$  0.04 & $-0.03~\pm$  0.05 & ITER \\
HIP 83120 & 5072 &  4.51 & $+0.25$ & $+0.23~\pm$  0.05 & $+0.33~\pm$  0.05 & ITER \\
HIP 87182 & 4687 &  4.63 & $+0.02$ & $-0.02~\pm$  0.03 & $-0.04~\pm$  0.03 & ITER \\
HIP 98535 & 5181 &  3.89 & $-0.18$ & $-0.17~\pm$  0.03 & $-0.23~\pm$  0.03 & ITER \\
HIP 102766 & 4988 &  4.60 & $-0.02$ & $-0.04~\pm$  0.05 & $-0.05~\pm$  0.03 & ITER \\
HIP 102582\tablenotemark{c} & 4574 &  4.72 & $-0.53$ & \nodata & \nodata & SME VESTA \\
 \enddata
 \tablenotetext{a}{Errors on \teff, log~$g$, and [Fe/H] stars are 44~K, 0.06~dex, and 0.03~dex for all stars.}
 \tablenotetext{b}{ITER: parameters determined using {\it Hipparcos} parallaxes and $Y^2$ isochrones. VESTA: parameters determined using classical SME fitting (no parallax information included) with a correction using Vesta as described in \citet{2005ApJS..159..141V}.}
 \tablenotetext{c}{The fit for HIP 102582 was run in ITER mode, but failed to converge. The initial fit suggests that the derived [Fe/H] is reliable, but other derived parameters are discarded for this star.}
 \label{tab:primaries}
 \end{deluxetable*}

The remaining 19 primaries are late-K or early-M dwarfs. For these targets we calculated their metallicities by applying the empirical methods from M13 to SpeX spectra. M13 provides empirical calibrations between observed atomic and molecular line strengths and the metallicities of stars from K5 to M5 for visible, $J-$, $H-$, and $K-$band spectra. We calculated the weighted mean of the $H-$ and $K-$band metallicities accounting for measurement (mostly Poisson) and calibration errors. For stars with visible wavelength spectra available from \citet{Lepine:2013lr} we included the visible calibration metallicities. Calibration errors (typically 0.08~dex) were {\it not} assumed to be uncorrelated because they are based on the same underlying sample and technique. Thus the calibration errors represent the error floor on the metallicities of these stars. Adopted metallicities for these stars are reported in Table~\ref{tab:sample}.

\subsection{Spectral Types}\label{sec:spt}
We used a custom by-eye matching routine to determine the spectral type of each companion or M dwarf primary. The routine uses NIR spectra and is based on the by-eye matching routine from the HAMMER spectral typing suite \citep{2007AJ....134.2398C}. Our routine separately displays the normalized $J$-, $H$-, and $K$-band spectrum from the target alongside a NIR template spectrum from \citet{Rayner:2003lr} or \citet{Cushing:2005lr}. An initial guess template is shown based on a $\chi^2$ comparison of the target and NIR templates. The user is allowed to switch templates (both spectral type and luminosity class) manually to get a better by-eye match. Half subtype templates were constructed (if they are not already included) from normalized, linear combinations of spectral pairs (e.g., an M5.5 is constructed by adding together an M5 and an M6). We also repeated this method using the entire $JHK$ spectrum as a crosscheck, and found differences were $\le1$ subtype in all cases.

To test the reliability of our spectral-typing method we analyzed a sample of stars with both visible and NIR data from \citet{Reid:1995lr}, \citet{West:2011fj}, \citet{Lepine:2013lr}, or this program. Spectral types for these targets were determined from their visible-wavelength spectra \citep[based on the system of][]{Kirkpatrick:1991kx}. We found no significant systematic offset between the optical spectral types and those determined from our by-eye matching of NIR spectra. Based on the scatter we estimated the errors on assigned spectral types to be $\simeq0.5$ for M4.5--M7.5 and $\simeq0.8$ for M7.5--M9. The higher errors for the latest spectral types may be in part due to systematic discrepancies between assigned spectral types in the optical from different surveys.

\citet{Rojas-Ayala:2012uq} and \citet{Newton:2014} presented relations between the empirical \water-K2 index and the spectral type of the star. The \citet{Newton:2014} is more relevant, as it includes more late-type stars. Interestingly we found that the \citet{Newton:2014} relation predicts spectral types systematically 0.5 subtypes later than those from our by-eye analysis. Since the spectral types from the \water-K2 index are based on the continuum shape, while our matching is based on the more traditional method of matching indices (albeit by eye), we only report our spectral types. This also kept the spectral types more consistent with the M13 study, which are based on optical spectra.

\section{Applicability of Prior Calibrations to late-M dwarfs}\label{sec:priorcal}
We examined the performance of previous M dwarf metallicity calibrations using our ultracool dwarf companion sample. Our sample covers a different range of spectral types than those used for previous calibrations, which focused on early- to mid-M dwarfs. The goal is to determine how these calibrations work (or fail) for the latest M dwarfs.

We followed the procedures for measuring metallicities given in \citet{Terrien:2012lr}, M13, and \citet{Newton:2014} to determine the metallicities of each companion. Each of these calibrations uses SpeX spectra to measure the equivalent widths of strong lines in the $H$- or $K$-band, although the set of lines varies. Thus applying their methods required using different feature (wavelength) definitions, as well as different procedures to estimate the (pseudo-)continuum for a given feature. The calibration of \citet{2010ApJ...720L.113R,Rojas-Ayala:2012uq} based on the TripleSpec spectrograph was not tested, because there are small but significant systematic offsets between equivalent widths using the different instruments \citep{Newton:2014} and because the targets from \citet{2010ApJ...720L.113R} were already folded into the \citet{Newton:2014} analysis.

\begin{figure*}[htbp] 
   \centering
   \includegraphics[width=\textwidth]{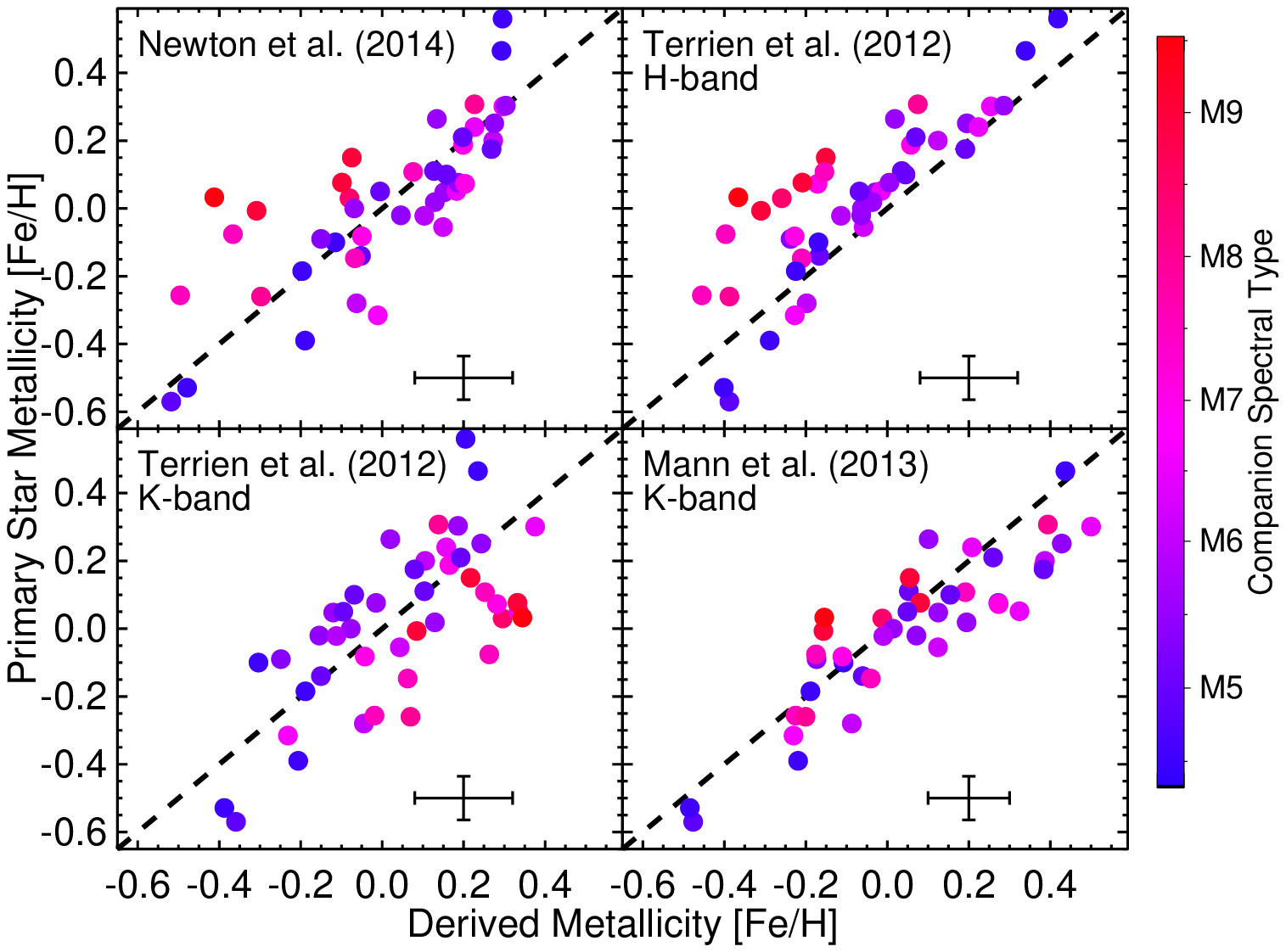} 
   \caption{Primary star (system) metallicity as a function of the metallicity determined for our ultracool companion sample based on the calibrations from \citet{Newton:2014}, \citet{Terrien:2012lr} $H$-band, \citet{Terrien:2012lr} $K$-band, and \citet{Mann:2013gf} $K$-band. Data points are colored by spectral type. The error bar in the bottom right denotes a typical (median) error from the primary stars and the error in the calibration reported from the relevant reference. The dashed line has slope unity and is added for reference. }
   \label{fig:cal_comp}
\end{figure*}

We show the derived companion metallicity using each of the literature calibrations versus the primary star metallicity in Figure~\ref{fig:cal_comp}. For comparison, we calculate the adjusted coefficient of determination (\rapsq) for each relation applied to our wide binary sample. \rapsq is defined as:
\begin{eqnarray}\label{eqn:rapsq}
R_{\mathrm{ap}}^2 = 1 - \frac{(n - 1)\sum(y_{i,\mathrm{model}}- {y_i})^2}{(n-p)\sum(y_i - \bar{y})^2},\\ \nonumber
\end{eqnarray}
where $p$ is the number of changeable parameters, $n$ is the number of data points in the fit, $y_i$ primary star metallicity of the $i$th star, $y_{i,model}$ is the metallicity of the $i$th star predicted by the fit, and $\bar{y}$ is the average of $y$. A R$_{\mathrm{ap}}^2$ closer to 1 implies that the model accurately explains the variance of the sample whereas R$_{\mathrm{ap}}^2$=0 implies that it can explain none. We report \rapsq\ and the standard deviation ($\sigma$) for each relation in Table~\ref{tab:stats}.

\begin{deluxetable}{lcccccc}
\tablecaption{Tests of Each Calibration on Mid to Late-M Dwarf Sample}
\tablehead{
\colhead{Reference} & \colhead{Band}& \colhead{\rapsq} & \colhead{$\sigma$} \\
}
\startdata
\citet{Newton:2014} & $K$ &  0.56 &  0.15\\
\citet{Terrien:2012lr} & $H$ &  0.47 &  0.13\\
\citet{Terrien:2012lr} & $K$ &  0.37 &  0.18\\
\citet{Mann:2013gf} & $K$ &  0.62 &  0.12\\
This work & $K$ & 0.89 & 0.07
\enddata
\label{tab:stats}
\end{deluxetable}

All prior calibrations show significant systematics with spectral type, resulting in inaccurate metallicities for the latest-type dwarfs. This is expected, since these calibrations were based almost entirely on stars $\simeq$M5 and earlier. If we remove the stars later than M6 from the sample, all calibrations yield results (as determined by \rapsq) consistent with those reported in the respective paper. 

The M13 calibration shows the least systematics with spectral type, in that this calibration accurately predicts the metallicity of M7-M9 dwarfs but on average underestimates the metallicities of the whole sample. The calibration from \citet{Newton:2014} performs reasonably well on our sample, most likely because their sample includes more mid-M dwarfs than other analyses. However \citet{Newton:2014} assigns incorrect metallicities for stars with [Fe/H]$>+0.3$ (already noted by Newton et al. 2014) and underestimates the metallicities of stars M7 and later. The two calibrations from \citet{Terrien:2012lr} have a similar issue with M7 and later stars, although the $K-$band calibration under- rather than overestimates the metallicity of the latest M dwarfs. 

\section{Measuring the Metallicities of Late-M Dwarfs}\label{sec:newcal}
The random and systematic errors present when we applied prior calibrations to the ultracool dwarf sample motivated a new metallicity calibration tuned for late-M dwarfs. To do this, we first used the list of metal-sensitive spectroscopic features and pseudo-continuum definitions from M13. We followed the method of M13 to determine which combination of features from this list gave the best calibration (as determined by \rchisq) for each wavelength regime ($J$, $H$, and $K$ band). We fit for calibrations of the form:
\begin{eqnarray}\label{eqn:fit}
\mathrm{[Fe/H]} &=& \sum_n (A_n(\mathrm{F_n}) + B_n(\mathrm{F_n})^2 )+\\ \nonumber &&C\mathrm{(H_2O-K2)} + D,
\end{eqnarray}
where $\mathrm{(H_2O-K2)}$ is a temperature-sensitive \water index defined by \citet{Rojas-Ayala:2012uq}, $F_n$ is the equivalent width of the $n$-th feature from M13, [Fe/H] is the metallicity of the system (derived from the primary star), and the other variables ($A$, $B$, $C$, $D$) were determined by Levenberg-Marquart least-squares minimization \citep{2009ASPC..411..251M}. 

Starting with one feature and no square term ($B_n=0$ and n=1) we tried all features from M13 for a given wavelength regime. After the best single feature was found we tried adding an additional feature. The total number of included features ($n$) was determined by an $F$-test, which measures whether the coefficient for the new term is consistent with zero. Additional terms are only added if the probability that the new coefficient is significant exceeds 95.5\% ($2\sigma$). The same test was applied to determine if squared terms should be included. Higher-order ($\ge3$rd) terms were not explored, but as we explain below, even second-order terms were not required.

Errors in equivalent widths and the \water-K2 index were calculated via Monte Carlo (MC). Noise was added to the spectrum equal to the measurement error computed by \texttt{Spextool}. Equivalent widths were then recalculated on the perturbed spectrum. This process was repeated 1000 times, and the error in a given $F_n$ and \water-K2 index was taken as the standard deviation of these 1000 values.

\begin{figure*}[htbp]
   \centering
   \includegraphics[width=\textwidth]{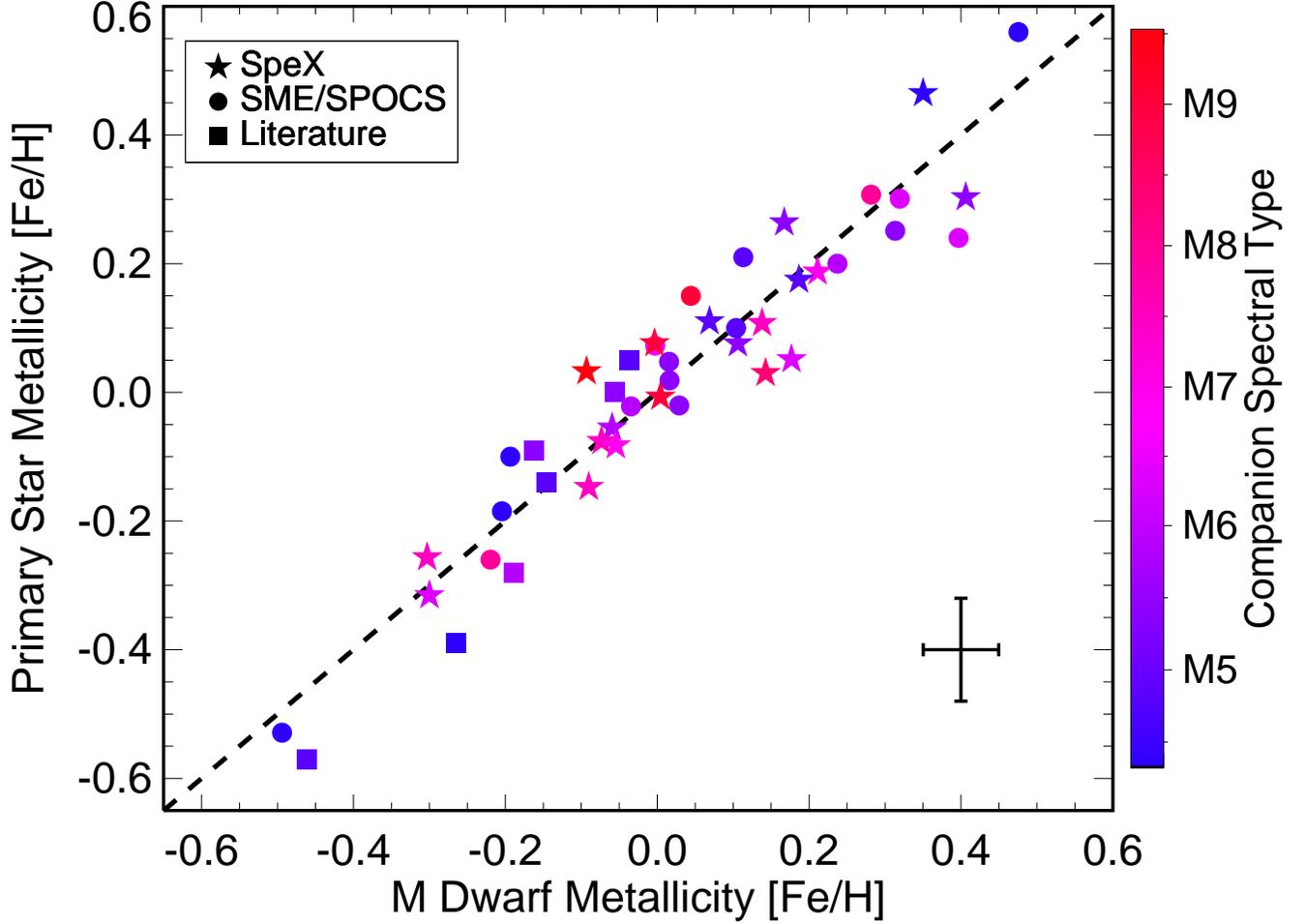} 
   \caption{Primary star (system) metallicity as a function of the metallicity derived for the ultracool companion sample based on Equation~\ref{eqn:cal}. Points are colored according to spectral type, and symbols indicate the source for the primary star metallicity. The error bar in the bottom right denotes the median error on primary star metallicity (y-axis) and the median measurement (mostly Poisson) error from estimating the M dwarf metallicity. The calibration error ($\simeq0.07$~dex) should be considered separately and is not shown.  }
   \label{fig:calibration}
\end{figure*}

We found the best fit in the $K$-band required just the Na~I and Ca~I lines:
\begin{eqnarray}
\label{eqn:cal}
\mathrm{[Fe/H]} &=& 0.131(\mathrm{EW_{Na}}) + 0.210(\mathrm{EW_{Ca}}) - \\ \nonumber &&3.07\mathrm{(H_2O-K2)} + 1.341,
\end{eqnarray}
where equivalent widths (EW$_{\mathrm{Na}}$ and EW$_{\mathrm{Ca}}$) are given in \AA \footnote{An IDL program for applying Equation~\ref{eqn:cal} can be found at \href{http://github.com/awmann/metal}{http://github.com/awmann/metal}}.
 We show the binary system metallicities as a function of the derived metallicity in Figure~\ref{fig:calibration}. Equation~\ref{eqn:cal} yielded a scatter ($\sigma$) of 0.07~dex, a \rapsq\ of 0.89, and a \rchisq\ of 2.1 ($\nu=39$). The quality of the calibration is similar to that of M13 for K5--M5 dwarfs, and significantly better than applying previous calibrations developed for early--mid M dwarfs (Section~\ref{sec:priorcal}). Feature and continuum regions for all measured spectral lines are identical to those in M13, but the two main features used in this work (Ca~I and Na~I) are also listed in Table~\ref{tab:lines}.

\begin{deluxetable}{l | c | c | c }
\tablewidth{0pt}
\tablecaption{Spectral features used}
\label{tab:lines}
\tablehead{\colhead{Name} & \colhead{Feature} & \colhead{Blue continuum} & \colhead{Red continuum}\\ 
\colhead{} & \colhead{($\micron$)} & \colhead{($\micron$)} & \colhead{($\micron$)}}
\startdata\label{tab:lines}
Na~I & 2.2045 -- 2.2113 & 2.1940 -- 2.1985 & 2.2130 -- 2.2190  \\
Ca~I & 2.2610 -- 2.2670 & 2.2450 -- 2.2520 & 2.2717 -- 2.2781  
\enddata 
\label{tab:lines}
\end{deluxetable}

\begin{figure}[htbp] 
   \centering
   \includegraphics[width=0.45\textwidth]{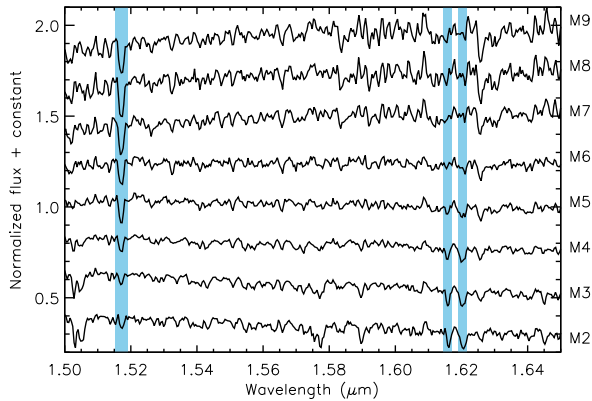} 
   \caption{$H$-band spectra of M dwarfs with a range of spectral types. The metal-sensitive features identified by \citet{Terrien:2012lr} are shown in teal. The Ca~I lines near 1.62~\um\ are strong and easily measured out to M4, but become impractical weak for later spectral types. The K~I line blueward of 1.52~\um\ is easy to measure for the full range, but has a strong spectral type dependence that is difficult to remove. }
   \label{fig:hband}
\end{figure}

We attempted to find a metallicity relation useful for $J-$ and $H$-band spectra following an identical prescription for the $K$-band. However, we found that calibrations with comparable performance (similar \rchisq, $\sigma$, and \rapsq) to Equation~\ref{eqn:cal} required using three or more features and still showed significant systematics with metallicity. We were able to find a formula with a relatively low $\sigma$ (0.12~dex), but such calibrations systematically underestimated the metallicity of the most metal-rich stars and systematically overestimated the metallicity of the most metal-poor stars. The primary cause is that the most effective features for measuring metallicity for K5--M5 become weak and difficult to measure past M4 (Figure~\ref{fig:hband}). The situation is even worse in the $J$-band where there were similar issues measuring features but we were also plagued by much lower S/N.

\citet{Newton:2014} found that the best $K$-band calibration was achieved using just the Na~I index, including a square term, and without the \water-K2 index. In contrast, we found that the inclusion of any squared term ($B_n$) was not justified by an $F$-test, but that the inclusion of the \water-K2 index was well justified by the same test. The \citet{Newton:2014} sample is calibrated on a relatively small range of spectral types (mostly M3--M5), thus it was probably not necessary to include the \water-K2 index, which was designed to help adjust for spectral changes with \teff{} \citep{Rojas-Ayala:2012uq}. Like \citet{Newton:2014} we find a second-order term is justified if we use just one feature (just Na~I). However, we find we get better results using Na~I and Ca~I than using higher order terms.

\begin{figure}[htbp] 
   \centering
   \includegraphics[width=0.45\textwidth]{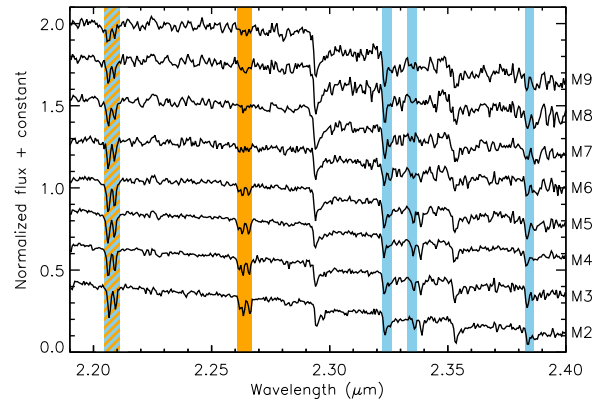} 
   \caption{$K$-band spectra of M dwarfs with a range of spectral types. Features used in Equation~\ref{eqn:cal} are shown in orange, with features used in M13 marked in teal. The labeled features at 2.345~\um\ and 2.385~\um\ are useful metallicity indicator to $\simeq$M5, after which they becomes difficult to measure.}
   \label{fig:kband}
\end{figure}

M13 found a best fit relation using multiple Na~I and CO bands in the $K$-band, although they also identified the Ca~I triplet as a strong metallicity indicator. However, the Na~I line at $\simeq$2.335~\um\ weakens past M5, while the Na~I doublet at 2.208~\um\ is only mildly sensitive to spectral type (Figure~\ref{fig:kband}). Here we found that the CO bands past 2.28~\um, despite being very strong, are poor indicators of metallicity for M5--M9, and thus are not useful in this calibration. The Ca~I triplet at 2.265~\um\ does become weaker for the coolest stars, but is still measurable even in the M9 dwarfs (Figure~\ref{fig:kband}) and remains a reliable predictor of metallicity based on our analysis.

Interestingly, the Na~I and Ca~I lines used in our calibration are the same two features identified by \citet{2010ApJ...720L.113R} to determine the metallicity of early M dwarfs. Because our method of finding the best calibration considers all metal-sensitive features identified by M13 and has no preference for these particular lines, this is a strong verification for the power of these atomic lines to measure the metallicity of M dwarfs.

\section{The Robustness of Our Calibration}\label{sec:tests}
\subsection{Chance Calibration Probability}\label{sec:chance}
Feeding a large number (30) of features to Equation~\ref{eqn:fit} increases the possibility of getting a reasonable calibration simply by chance, since each feature adds several potential degrees of freedom. Finding a calibration of quality similar to that of Equation~\ref{eqn:cal} is unlikely. However, we want to better quantify {\it how unlikely} this is.

As a test, we reassigned the metallicities of each primary star randomly to another binary in the sample. We then reapplied our method as described above: feeding in metal-sensitive features from M13 into Equation~\ref{eqn:fit}, finding the fit by least squares, and adding in more features until the change in \rchisq\ is minimal. We then repeated this process 10,000 times, each time with re-randomized metallicities (although binaries are forbidden from having their original metallicity) and recording the \rapsq\ and rms values of the final fit. The rms is often a poor metric when comparing multiple fits, because it is sensitive to the underlying distribution. For example, if our primary star metallicities were clustered around $-0.2<$[Fe/H]$<+0.2$, a calibration that assigns all stars solar metallicity would have an rms $\ll0.2$, even though such a calibration would not be useful. This is why we reassigned, rather than randomized metallicities: to preserve the sample's metallicity distribution. 

We found that the 99.7\% highest (3$\sigma$) \rapsq\ value amongst our random sample is 0.41, which is less than half of that from our fit. Similarly, the 99.7\% lowest rms is 0.19, more than twice as high as the rms of Equation~\ref{eqn:cal}. Thus the probability of getting an \rapsq\ of 0.89 and an rms of 0.07 is $\ll 0.1\%$, demonstrating that our method is statistically significant despite the introduction of a large number of variables into the fit.

\subsection{Systematics}
We searched for systematic issues in our calibration by comparing the fit residuals with spectral type, metallicity, and source of metallicity for the primary using a Spearman rank test. For all three cases we found no statistically significant correlation (P=0.31, 0.13, 0.06, respectively), suggesting that our calibration is robust over the range of metallicities and spectral types covered in our sample. This was slightly complicated by the distribution of points for these three parameters. For example, our sample includes only four dwarfs M9 or later, all of which have near solar metallicity ($-0.05 <$ [Fe/H] $< +0.20$).

To test the limits of our calibration we applied Equation~\ref{eqn:cal} to two additional stars outside the range of our calibrators. HIP 114962B is an M3.5 subdwarf  companion to an F8 subgiant, and GJ 1048B is an early-L  dwarf companion to a K2 dwarf \citep{Gizis:2001}.  HIP 114962 has [Fe/H]$=-1.40\pm$0.08 \citep{2001A&A...373..159C,2011A&A...530A.138C,Lee:2011}, and GJ 1048 has [Fe/H]=$+0.06\pm0.03$ \citep{2012A&A...545A..32A}. These two pairs were not included in the initial calibration sample because they are outside the range of spectral types considered (see Section~\ref{sec:sample}). However, they are still useful tests because if our calibration fails just outside the range of companion star spectral types it suggests a problem.

Applying Equation~\ref{eqn:cal} to spectra of these two companions yielded metallicities of [Fe/H] = -1.26 and 0.13 for HIP 114962B and GJ 1048B, respectively. Accounting for measurement and calibration errors, the differences between derived and primary star metallicities were 1.2$\sigma$ and 0.7$\sigma$, respectively. The agreement suggests that the calibration may be effective slightly outside the range of spectral types of the calibrators.

\subsection{Unresolved Binaries}
Approximately $45\%$ of wide M dwarf binaries contain at least one more star \citep{2010ApJ...720.1727L}, typically a close ($<30$~AU separation) companion to one of the components (or a close companion to each of the components for quadruple systems). At the median distance to our targets (32~pc) any such close-in companion would be unresolvable (separations $\lesssim$ 1 arc second) in the SpeX slit-viewing camera. Unresolved companions to the primary star can be identified as a spectroscopic binary in our ESPaDOnS spectra. However, for both M dwarf primaries and all companions, the spectra was not high enough resolution to detect multiple lines. The presence of an unresolved star may change the \water-K2 index and continuum measurements, which will in turn add scatter to the calibration.

We tested the effect of binarity on our calibration and the M13 calibration using a sample of bright late-K and M dwarfs from LG11 or this program. More than 400 of these targets have NIR SpeX spectra, primarily from a program aimed at studying the properties of M dwarfs and their planets (Gaidos et al. in preparation). NIR data for these stars were taken, reduced, and analyzed with the exact same techniques used for this work (see Section~\ref{sec:obs}). We selected the 253 K7-M8 dwarfs with parallaxes from {\it Hipparcos} \citep{van-Leeuwen:2007yq} or ground-based surveys \citep{2005AJ....130..337C, 2006AJ....132.2360H, 2009AJ....137.4109L, 2011AJ....141..117J, 2014ApJ...784..156D}. We calculated the expected distance to each star using the mean of the \water-K2 -- $M_K$ relation from \citet{Newton:2014} and the $M_J$-spectral type relation from \citet{Lepine:2013lr}. We removed 65 dwarfs with spectroscopically determined distances $>3\sigma$ larger than those based on trigonometric parallax, as these are likely unresolved binaries.

We used this sample to construct 200 unique artificial binary spectra. Specifically, we randomly combined two stars with metallicity differences $<0.07$~dex (similar to the measurement error), which we determined from each spectrum using the calibration from M13 for K7-M4, and Equation~\ref{eqn:cal} for M4.5--M8. To accurately place the stars at the same distance, we normalized and scaled each spectrum according to their $M_K$. We calculated the masses of each star using the empirical relation from \citet{2000A&A...364..217D}. We assigned a random orbital period following the log-normal distribution from \citet{2010ApJS..190....1R}, but with a cutoff at semi-major axes of 32 AU. This cutoff corresponds to a 1\arcsec\ separation (resolvable in the SpeX guider) at the median distance to our targets. We then calculated the radial velocity shift assuming a random inclination and circular orbit, which we applied to the fainter of the two stars. We combined the two resulting spectra to form an empirical binary spectrum. We analyzed the resulting spectrum just as we would our other observations, determining the spectral type, moving the star to its rest frame, and recalculating the metallicity following the method as was applied for the single stars. 

\begin{figure}[htbp] 
   \centering
   \includegraphics[width=0.5\textwidth]{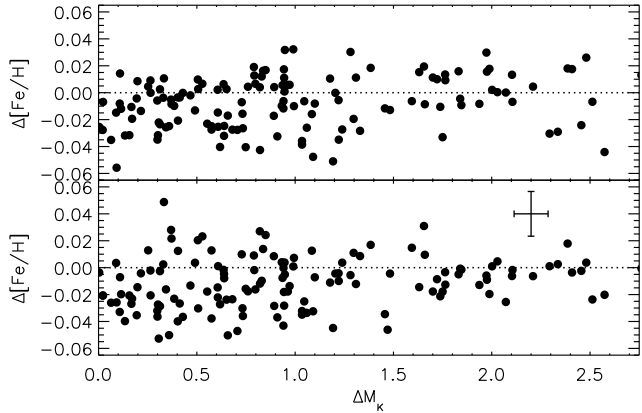} 
   \caption{Difference in metallicity (as determined by Equation~\ref{eqn:cal} or calibration of M13) of synthetic binary formed from combining the NIR spectra of two single K7-M8 dwarfs of similar metallicity. The top plot shows difference in metallicity between the synthetic binary and the mean of the two components, while the bottom plot shows the difference compared to the brighter of the two components. The error bar shows the median error in [Fe/H] (measurement error only) and $M_K$. The dotted line indicates zero difference.}
   \label{fig:binary}
\end{figure}

Figure~\ref{fig:binary} shows the difference between the metallicity derived for the artificial binaries and the mean metallicity of their components as a function of $\Delta M_K$. We found that there is a small bias in favor lower metallicities, but the median difference is only $-0.01$~dex. This offset arises due to large radial velocity variations smearing out the features resulting in slightly lower equivalent widths. However, even in cases of tight, similar mass binaries, this effect is small compared to calibration and measurement errors. The scatter in metallicities is only 0.02~dex and no pairs changed in metallicity by more than 0.06~dex. Most of the scatter can be explained by small ($<0.07$~dex) differences in the metallicity of the two components and the additional measurement noise. 

\subsection{M+M Wide Binaries}
The technique of calibrating metallicity diagnostics using wide binaries relies on the assumption that wide binary components have the same metallicity. This assumption must be at least partially valid, since we would not be able to derive such a precise calibration if the metallicities of the primary and companion were uncorrelated (see Section~\ref{sec:chance}). However, if instead there was a dispersion in the metallicity of the star-forming cloud, or one of the stars accreted metal-rich material over its lifetime, the binary elements could have similar but not identical metallicities.

Examination of FGK+FGK wide binary systems have found consistent metallicities to within expected measurement errors \citep[e.g.,][]{2004A&A...420..683D}, suggesting that metallicity differences between binary components, if present, are very small. \citet{Rojas-Ayala:2012uq} showed that, within errors, their method to measure M dwarf metallicities gave consistent results for both components of five M+M pairs. However, \citet{Rojas-Ayala:2012uq} found that the color-magnitude metallicity estimates \citep{Johnson:2009fk, Schlaufman:2010qy} did not show the same consistency. Thus it is prudent to apply a similar test using our calibration.

\begin{deluxetable*}{l c c l c c c c}
\tablewidth{0pt}
\tablecaption{M+M Binaries}
\tablehead{ \multicolumn{3}{c}{Primary} & \multicolumn{3}{c}{Companion} \\
\colhead{Name} & \colhead{SpT} &\colhead{[Fe/H]} &\colhead{Name} & \colhead{SpT} & \colhead{[Fe/H]} & \colhead{$\Delta$[Fe/H]$\pm\sigma$\tablenotemark{a}} } 
\startdata
\label{tab:mms}
GJ 1245A & M6.0 & $+0.02$ & GJ 1245C & M5.5 & $-0.01$ & 0.03$~\pm$~0.05\\
GJ 896A & M4.5 & $-0.07$ & GJ 896B & M4.5 & $-0.04$ & 0.03$~\pm$~0.03\\
GJ 118.2B & M0.0 & $+0.20$ & GJ 118.2C & M3.5 & $+0.25$ & 0.05$~\pm$~0.03\\
GJ 617A & M0.5 & $+0.14$ & GJ 617B & M2.5 & $+0.20$ & 0.06$~\pm$~0.02\\
GJ 4049A & M3.0 & $-0.17$ & GJ 4049B & M3.5 & $-0.18$ & 0.01$~\pm$~0.03\\
GJ 725A & M3.0 & $-0.29$ & GJ 725B & M3.5 & $-0.33$ & 0.04~$\pm$~0.01\\
LP 213-67 & M7.0 & $-0.01$ & LP 213-68 & M8.0 & $-0.02$ & 0.01~$\pm$~0.06
\enddata 
\tablenotetext{a}{Includes measurement error only.}
\label{tab:mms}
\end{deluxetable*}

We selected a sample of seven M+M wide binaries following the same methods as explained in Section~\ref{sec:sample}, with the restrictions that both components are earlier than M4.5, or both components are M4.5 or later (so the same calibration can be used) and that the components have separations $>5$\arcsec (so that each star can be studied separately). The sample is listed in Table~\ref{tab:mms}.

We measured the metallicities of each component of these wide pairs, using the M13 calibration for M0--M4 dwarfs, and Equation~\ref{eqn:cal} for M4.5--M9 dwarfs. We found a median difference between primary and companion metallicity of 0.01~dex and a maximum difference of 0.06~dex. Most of these differences were similar in size to the measurement errors, and all differences were less than the calibration errors. 

\subsection{Bridging the Calibrations}
Combining our work with that of M13 it should be possible to measure the metallicities of dwarfs from K5 to M9.5. Although since these methods are calibrated on a different set of stars there is a possibility there will be systematic differences between the two calibrations. To investigate this, we applied Equation~\ref{eqn:cal} and the $K-$band calibration from M13 to a sample of 15 M4-M5 stars from \citet{Lepine:2013lr}. The metallicities between the two calibrations for 14 of these stars were consistent within 1$\sigma$, with the remaining star showing a difference of 1.6$\sigma$. We found no evidence of a systematic offset between the two sets of derived metallicities for these stars (median difference = 0.02~dex). 

\section{Summary}\label{sec:discussion}
We have used wide binaries containing an F, G, K, or early M dwarf primary with a M4.5--M9.5 companion to calibrate spectroscopic metallicity diagnostics for the coolest M dwarfs. Although many calibrations already exist, based either on spectroscopy or absolute magnitude, none have been calibrated with the latest M-type dwarfs. We showed that these prior spectroscopic calibrations yield systematically inaccurate metallicities for the coolest M dwarfs (Figure~\ref{fig:cal_comp}). We derived a new calibration for late-M (M4.5-M9.5) dwarfs and found that the Na~I doublet and Ca~I triplet were the most effective metallicity indicators for late-M dwarfs. We found that our calibration (Equation~\ref{eqn:cal}) predicts metallicities accurate to $\simeq0.07$~dex for $-0.58<$[Fe/H]$<+0.56$. The error is comparable to that reported by M13 for the K5--M5 sample. By combining this work with that of M13 ,it is possible to measure metallicities of stars across the entire M dwarf sequence.

For the F, G, and K dwarf primaries [Fe/H] is generally measured directly using the plethora of Fe lines present in their spectra. However, measuring metallicities of M dwarfs generally relies on Ca and Na. We would therefore expect to get a smaller scatter relating the strength of these features to [$\alpha$/H], [M/H], and [Na/H]. Unsurprisingly, this was seen in previous studies using similar features \citep[e.g.][]{Rojas-Ayala:2012uq,Mann:2013gf}. This also may be the source of the higher scatter between primary and companion metallicity seen at lower metallicity (M13). However, most of the literature sources we use only report [Fe/H]. Although calibrations exist to determine [M/H] for early M dwarfs, these calibrations use almost identical lines to the [Fe/H] calibrations, which may complicate the result. Thus we would be left with only 18 stars, which is not enough for a meaningful investigation. Future analysis of this issue would require a more homogenous analysis of the primary stars.

We performed a number of tests to assess the quality and applicability of the calibration. We verified that:
\begin{itemize}
\item Despite the use of a large line list and many free parameters, the precision of the calibration cannot be due to chance ($P\ll0.001$).
\item The metallicities predicted by Equation~\ref{eqn:cal} are free of significant trends as a function of spectral type, metallicity, or source of metallicity for the primary.
\item The calibration (and that of M13) is unaffected by unresolved binaries (triples) in the calibration sample. 
\item Both this calibration and that of M13 yield consistent metallicities for each component of M+M wide binaries. 
\item The calibration from M13 and this work predict metallicities for M4-M5 dwarfs (where the calibration samples overlap) that are in agreement.
\end{itemize}

Another potential source of error is the presence of false common-proper-motion companions (chance alignment) in the calibration sample. Presumably two unassociated stars will have random metallicities, and therefore appear as outliers in our relation. The lack of outliers in Figure~\ref{fig:calibration} suggests our sample is relatively free of false binaries.  Based on the published proper motions and statistical arguments from \citet{Lepine:2007qy} and \citet{Tokovinin:2012fj} we expect the false-binary rate to be $\ll8\%$. The true number is probably significantly lower than this, as many pairs from the literature are identified using distance and radial velocity information in conjunction with proper motions. 

Although the Na~I and Ca~I lines are strong metallicity indicators for M dwarfs, the Na~I doublet becomes significantly weaker and the Ca~I triplet is essentially not detectible in L dwarfs at this S/N and resolution. It is promising that our current calibration works for a single L dwarf, however it is hard to draw conclusions from a single star. Extending this calibration further into the L dwarf regime may require fine-tuning the calibration and/or using an entirely different set of lines. We will investigate measuring the metallicities of L dwarfs in a future paper.

\acknowledgments
We thank the anonymous referee for their useful comments on this paper. This work was supported by the Harlan J. Smith Fellowship from the University of Texas at Austin to AWM, NASA grants NNX10AI90G (Astrobiology: Exobiology \& Evolutionary Biology) and NNX11AC33G (Origins of Solar Systems) to EG; and NSF grant AST09-09222 to MCL, 0822443 to KMA, and AST-0709460 to EAM. Based on observations obtained at the Canada-France-Hawaii Telescope (CFHT) which is operated by the National Research Council of Canada, the Institut National des Sciences de l'Univers of the Centre National de la Recherche Scientifique of France, and the University of Hawaii. Also based on observations obtained with the Infrared Telescope Facility, which is operated by the University of Hawaii under Cooperative Agreement no. NNX-08AE38A with the National Aeronautics and Space Administration, Science Mission Directorate, Planetary Astronomy Program. This research has made use of the Washington Double Star Catalog maintained at the U.S. Naval Observatory. The authors wish to recognize and acknowledge the very significant cultural role and reverence that the summit of Mauna Kea has always had within the indigenous Hawaiian community. We are most fortunate to have the opportunity to conduct observations from this mountain.

{\it Facility:} \facility{IRTF:SpeX}, \facility{CFHT:ESPaDOnS}

\bibliography{$HOME/dropbox/fullbiblio.bib}


\clearpage

\end{document}